# Analogy Search Engine: Finding Analogies in Cross-Domain Research Papers

**Jieli Zhou, Yuntao Zhou, Yi Xu**

**Advisors: Joel Chan, Niki Kittur, John Stamper**
Language Technologies Institute
Carnegie Mellon University
Pittsburgh, PA

## Abstract

In recent years, with the rapid proliferation of research publications in the field of Artificial Intelligence, it is becoming increasingly difficult for researchers to effectively keep up with all the latest research in one's own domains. However, history has shown that scientific breakthroughs often come from collaborations of researchers from different domains. Traditional search algorithms like Lexical search, which look for literal matches or synonyms and variants of the query words, are not effective for discovering cross-domain research papers and meeting the needs of researchers in this age of information overflow. In this paper, we developed and tested an innovative semantic search engine, Analogy Search Engine (ASE), for 2000 AI research paper abstracts across domains like Language Technologies, Robotics, Machine Learning, Computational Biology, Human Computer Interactions, etc. ASE combines recent theories and methods from Computational Analogy and Natural Language Processing to go beyond keyword-based lexical search and discover the deeper analogical relationships among research paper abstracts. We experimentally show that ASE is capable of finding more interesting and useful research papers than baseline elasticsearch. Furthermore, we believe that the methods used in ASE go beyond academic paper and will benefit many other document search tasks.

## 1 Introduction

Over the last 20 years, number of publications in the field of Artificial Intelligence has grown continuously. The number has surged around the year 2012 (Figure 1), when Deep Learning techniques were first successfully used for difficult tasks like image classification [1], speech recognition [2]. Other AI techniques like Deep Reinforcement Learning have also gained a lot of popularity since AlphaGo beated the world Go Champion in 2016 in the game of Go[3] as witnessed by the number of research papers published under the topic of "Deep Reinforcement Learning" over the years (Figure 2). Nowadays in the age of information overflow, numerous science and technology blogs are out there for quick information digest and famous researchers constantly tweet about their latest work, but research papers still remain the most faithful and credible sources of research information. With over 3500 AI papers published in 2018 alone and more papers in this field to come in the future, it is becoming increasingly difficult for researchers to keep up with the latest research in one own fields, let alone getting informed of other relevant cross-domain research.

Current academic search engines search for exact-match or synonyms and variants of the query term. However, this poses two levels of insufficiencies. **First, researchers may not know what exact key phrases to use as query term.** For example, if one is interested in the applications of an algorithm X in the field of Y, by just searching the combinations of names, algorithm X and the field name Y on the current search engines will most likely get papers whose titles or full text have such keywords combinations match. There is no guarantee such keyword matches will lead to papers which actually uses algorithm X in field Y.

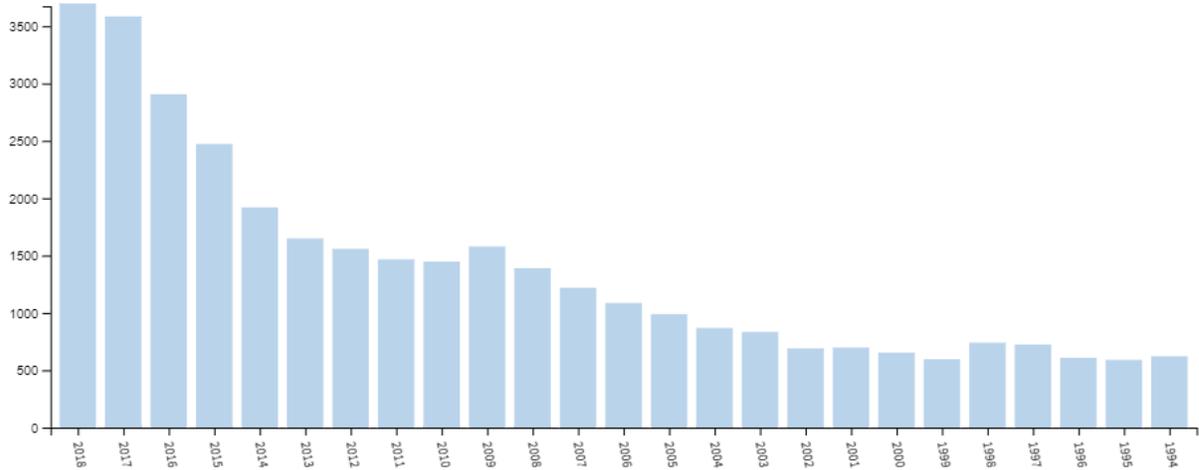

**Figure 1: Number of Artificial Intelligence Papers 1994-2018 (Source: Web of Science)**

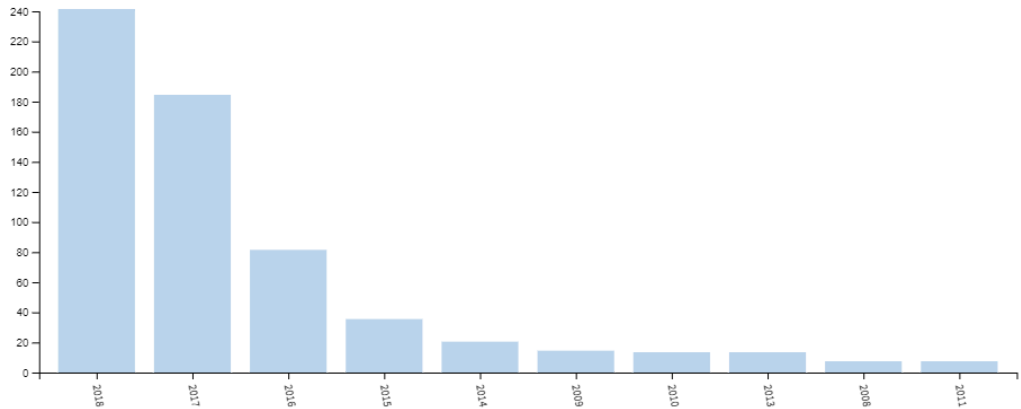

**Figure 2: Number of Papers with topic "Deep Reinforcement Learning" over the years (Source: Web of Science)**

**Second, researchers usually do not start with keywords for searching research papers, instead they usually start with some paper in mind, and want to explore other papers which share certain aspects with this paper.** However, current academic search engines capture more of the shallow word-level information than the deeper semantic meanings, which makes them ineffective for searching and recommending similar research papers.

With the recent development in computational analogy theory [4] and deep learning based distributed word representation, we want to see how these techniques can enable a better search engine for research paper abstracts.

## 2 Hypothesis and Research Goals

We make the hypothesis that the abstract of most AI research papers can be divided into 4 functional units, **Background** (the specific fields of the research), **Purpose** (the problem the research aims to solve), **Method** (the algorithm the research introduces to meet its purpose) and **Results** (the findings or algorithmic improvement the research gets). We also make the assumption that papers with similar purpose but different methods or with similar methods but different purposes may be innovative for researchers. This assumption is based on the successful findings in [4] where seeing product descriptions "close" in the purpose domain but "far" in the method domain are great for ideating innovative new products. We believe that researchers would benefit from this as well, since finding different methods for the same purpose could create new research opportunities for the same problem while finding different purposes for the same mechanism could enable new applications of the method.

In this paper, we focused on the first technique, finding papers with similar purpose but different mechanisms, but the second technique would be easy to implement as well (as explained in more detail in section 5). More specifically, we want to address three problems. 1. For a given paper abstract, can we classify it into different functional units, Background/Purpose/Method/Results? 2.With the functional units labeled for a given paper abstract, along with a corpus of other labeled paper abstracts, can we find interesting and new paper abstracts using text clustering and other NLP techniques? 3.For a given target AI paper, can we find more diverse and analogically relevant papers using our Analogy Search algorithms than baseline Information Retrieval models? The first problem is tested in [5] and we got 2K labeled paper abstracts from the authors of [5] using the same approach. We build a tool using this dataset and conduct user studies experiments to address the second and third problems.

## 3 Relationship to Prior Work

This work is inspired by the division of product description text into Purpose and Mechanism as in [4], the annotation of research paper abstracts as in [5] and the structured abstracts format as in PubMed[1]. We believe that the philosophy of dividing unstructured paper abstract into sequential and structured functional units (Background, Purpose, Method, Results) corresponds to how researchers both conduct and read research papers. The method of finding textual data similar in one aspect but different in another aspect is a follow-up application of the computational analogy theory as developed in [4]. Numerous works in computational analogy have shown that cross-domain analogy can produce remarkable scientific breakthroughs. Greek philosopher Chrysippus speculate that sound was a wave phenomena by finding analogy in observing water. Analogy to bicycle allowed the Wright brothers to design a steerable aircraft. We continue this line of theory and want to test whether it also applies to research papers and eventually helps researchers make such breakthroughs.

## 4 Spring/Fall Development Goals

### 4.1 Spring Landmarks

We began our project since February 2018 and by May 2018, we have cleaned out a paper corpus contained about 2000 papers with the abstracts labeled by crowdworkers into the above mentioned 4 categories, Background, Purpose, Method, Results. We ran the GloVe word embedding algorithm [6] to vectorize the text data in each functional units. We then developed three variants of Analogy Search algorithms (naive cosine, KNN-KMeans, Max-Min). We also made a demo search engine interface which supports ASE algorithms (Figure 3). Finally, we conducted an empirical user evaluation with five CMU Master students (Figure 4) to evaluate the effectiveness of token-level embedding for certain aspects of the abstract[2].

---

[1] https://www.nlm.nih.gov/bsd/policy/structured_abstracts.html

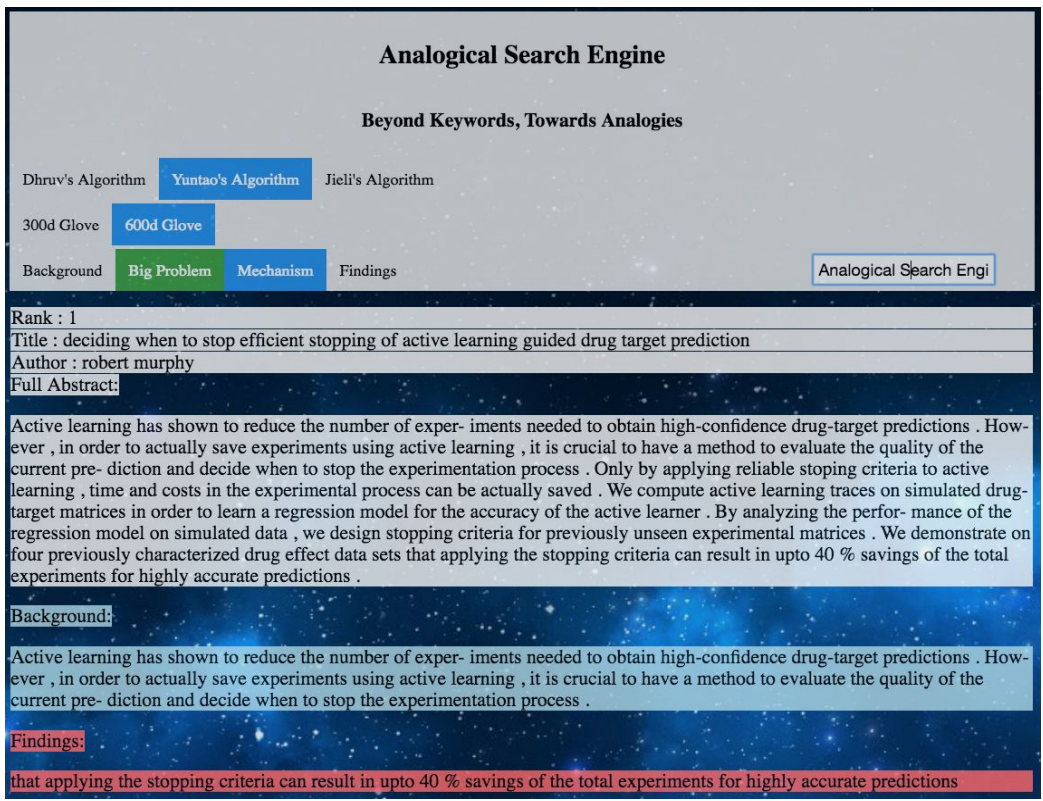

Figure 3: Demo Interface of Analogy Search Engine

The evaluation process is as follows.

1. We run our algorithm for a given paper using the given metric and rank all papers in our corpus.
2. We present 40 results for each paper. Among these result papers, 20 of them show up from the top of the returned list, i.e. (most relevant under the metric) while 20 of them are from the bottom of the returned results (least relevant under the metric).
3. We want the user to tell us by examining the corresponding segmented parts of the result paper abstract, which results are from the top 20, and which results are from the bottom 20.
4. By collecting feedbacks from users, we can test how effective is our ranking algorithm and how effective are the segmentation and vectorization of paper abstracts.

---

[2] https://zhoujieli.com/ase-evaluation

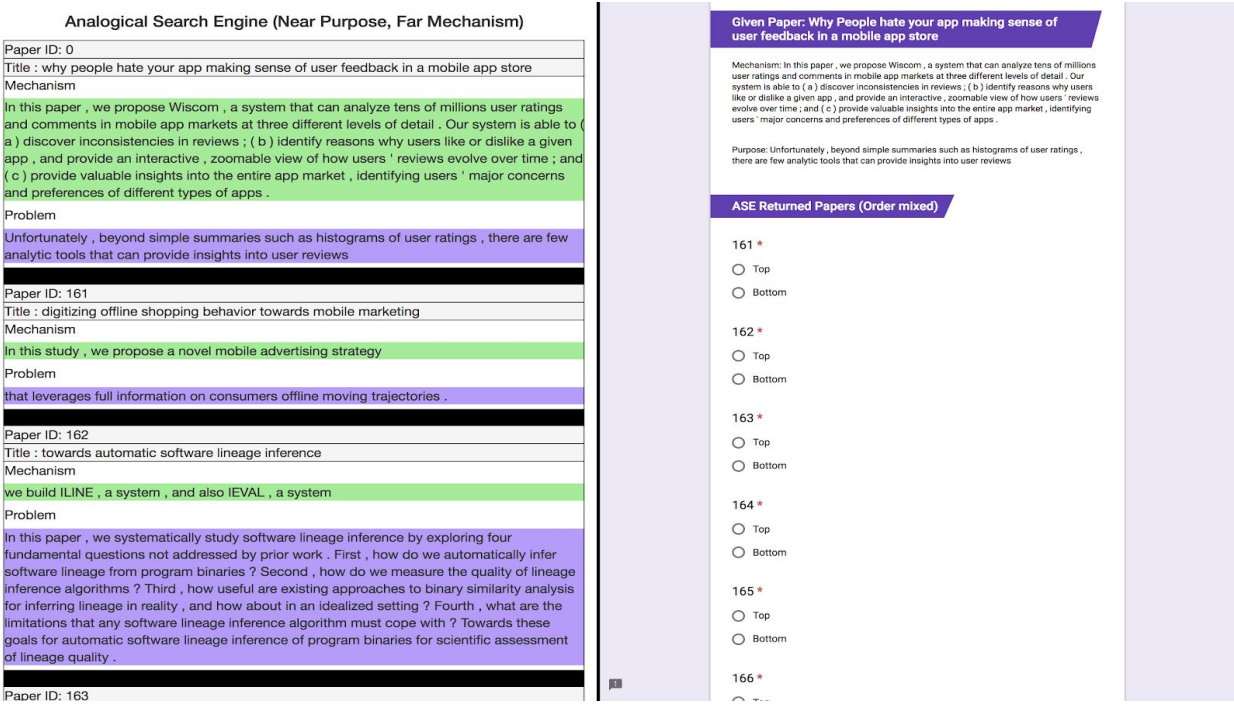

**Figure 4: Token-level embedding effectiveness user studies**

To quantitatively test how effective the tokens are represented by embedding, we devise the token effectiveness score (TES), that is after N users gave us feedbacks, we compare the majority vote (whether the users think a paper should appear on the top/bottom of the returned list as in step 2 above), if the majority vote indeed indicates the right position of the paper in the returned list, we mark that as a Correct Majority Vote, otherwise Incorrect Majority Vote. Then TES is computed as follows, where α is the Number of Correct Majority Votes, β is the Number of Incorrect Majority Votes, N is the total number of search results (40 in our case). Based on our definition, TES ranges from -1 to 1, where 1 is perfectly effective and -1 is totally ineffective. We report our TES results in the table below. Purpose and Findings units of the abstracts are most effectively represented.

$$TES = \frac{\alpha - \beta}{N}$$

| Functional Units | Tokens Effectiveness Score | #Correct Majority Votes ($\alpha$) | #Wrong Majority Votes ($\beta$) |
|---|---|---|---|
| Background | 0.2 | 24 | 16 |
| Purpose | 0.55 | 31 | 9 |
| Method | 0.25 | 25 | 15 |
| Findings | 0.85 | 37 | 3 |
| Near-Purpose, Far-Mechanism | 0.4 | 28 | 12 |

**Table 1: Token Effectiveness Score for different functional units in abstracts**

### 4.2 Fall Goals

We continued this project from September 2018 and our main goal is to quantitatively test whether Analogy Search works better than baseline information retrieval. Specifically, we interleave the top n search results of Elasticsearch and Analogy Search at random, enable the frontend functionality where the users could mark useful or not useful for each interleaved search result, then the backend would collecte and aggregate the user feedbacks and measure the overall usefulness and interestingness of the results returned by Analogy Search and Elasticsearch correspondingly. The details are described in section 7.

## 5. Dataset

The dataset we used to build ASE are the 300-d GloVe embeddings from the Solvent 2K dataset [5]. The dataset has the 2071 papers' abstract text segmented into 5 functional categories (background, problems, mechanism, method, findings) (Figure 5) and are transformed into 7 GloVe word embedding vectors (full_Abstract, background, big_problems/purpose, findings, mechanism, method, problem) for each paper (Figure 6). The ASE system is supposed to utilize these 300d vectors to search for papers that "near" on one aspect(e.g. purpose) and "far" on another aspect(e.g. mechanism). All these vectors are rescaled and normalized when applied to the ML models in the algorithm.

**Figure 5: token-level functional units segmentation for paper abstracts in the Solvent 2K dataset**

|  | 2017_1 | 2017_10 | 2017_11 | 2017_4 | 2017_5 | 2017 |
|---|---|---|---|---|---|---|
| abstract_tokens | [0.43347350111689603, -0.49879975944804705, -0... | [0.5662095452419671, 0.502941070278593, 0.4854... | [0.49234163403711406, 1.104922934249234, 0.289... | [0.48509416410259504, -0.066205609494076, -0.2... | [1.104698491235835, -0.3416858983046, 0.031509... | [-0.01868127075096000 -0.59728728593840( ( |
| background_tokens | [0.27222693695627603, -0.14244222103050302, 0.... | [0.43205178536014105, -0.22308430190296802, -0... | NaN | [0.36670372644584404, 0.030234543565292003, -0... | [0.833456078197855, -0.7185061123066551, 0.009... | [-0.430599543905568( -0.82412750950737 0.2 |
| big_problem_tokens | [0.375188582381668, -0.23917123075276, 0.12583... | [0.20582389192084302, -0.05855840756371301, 0.... | [0.7663105373478281, 0.631422280830779, -0.154... | [0.29640790517960003, -0.13793831956195102, 0.... | [0.51349227604865, -0.559687926568602, -0.0051... | [0.69401093804792 -0.3148045644478( -0.2954 |
| findings_tokens | [0.364476961339088, -0.823811780887003, -0.976... | [0.75500908849695, 0.7764743351094281, 0.31436... | [-0.06341022830863001, -0.005895438779612, -0.... | [0.36234283470850603, -0.562218101074425, -0.3... | [1.036201226204884, 0.09903122277705001, 0.220... | [-0.08963613568128 -1.25962178556349( -0.2 |
| mechanism_tokens | NaN | NaN | [0.39933824368326604, 0.250490358093143, -0.23... | NaN | NaN | Na |
| method_tokens | [-0.622416003436532, -0.349569300784692, 0.039... | [-0.30553842992621405, 0.758318664984948, 1.19... | [-0.062787393511986, 0.9891253085782401, -0.05... | [0.20040196454778503, -0.17392039097644402, -0... | [-0.068975810289891, -0.237562673864721020, -0... | [0.2773115956286 0.0182831375115930( -0.23 |
| problem_tokens | [0.556259062267702, -0.409280865540177030, -0.7... | [-0.237582779220179, 0.26391234534122604, 0.43... | [0.7663105373478281, 0.631422280830779, -0.154... | [0.25263880891948504, -0.242649724905330020, 0... | [-0.023589820416085, -0.29310025765115405, -0.... | [1.29957042833057 -0.04055374792475100 -0 |

7 rows × 2071 columns

Figure 6: 300d GloVe vectors for each functional units in the paper abstract of the Solvent 2K dataset

## 6. System Algorithm Design

We intend to design the algorithms for analogical search engine (ASE) such that users can find papers which are 1) highly relevant to their interesting query paper's purpose, and 2) are innovative and interesting enough to generate new methodology ideas for their future research. Accordingly, four algorithms are designed and implemented as part of the backend of ASE, i.e. Naive Cosine Distance, the KNN-KMeans, the Naive-Farthest and Farthest-Neighbor algorithms.

### 6.1 Algorithm Details

The Naive Cosine Distance Algorithm takes two steps to implement the logic of being "near" on one aspect and "far" on another. The system ranks the papers to the query paper on both "near" and "far" aspects orderly by applying cosine distance to the query paper. Basically it chooses the top 50 nearest paper to seed paper on "near" aspect, and rerank the 50 papers by the cosine distance (the "farthest" ones will be the top ones) to query paper on the "far" aspects and pick the top papers from the new list.

The KNN-KMeans algorithm has two core steps to pick the best analogically related papers. First step is to rank all the papers corpus to find the top nearest neighbors to the query paper by comparing the embedding vector of user's prefered "near" aspect(s). If multiple "near" aspects are chosen by user, the algorithm will combine each result list of "near" aspects and make the results weighted to pick the top related paper among all aspects. Then the system will cluster the whole corpus of 2k papers on "far" aspects by KMeans ( k = 20, which is determined by elbow evaluation of the dataset). With the clustering information on "far" aspects, the system will remove the papers that share the same cluster with the query paper. Finally the rest of the survivors in the candidates paper list will be the analogical search result. The logic here is that, the first step ensures papers in the potential candidates paper pool are closely related to the query paper according to the feature vector in "near" aspects. With the "near" paper pool, the second step is to remove the papers that are very similar to the query paper in "far" aspects from the candidate paper list, to avoid the case that some similar papers may appear to be less interesting to search engine users.

The Naive-Farthest and Farthest-Neighbor algorithm designs are both inspired by the MAX-MIN and MAX-AVG as in [4]. But instead of applying max-min and max-avg to ensure the diversity of result, the Naive-Farthest algorithm applies naive cosine similarity to rerank the list from the "Farthest" to the "Nearest", and Farthest Neighbor algorithm refers to a faster and more scalable dynamically-search algorithm to diverse the search result [7]. Both algorithms are implemented according to the assumption that user wants to find papers that is "near" in purpose and "far" in mechanism as described in section 2. Each algorithm has three steps. First, the whole paper corpus will be clustered by the purpose vectors information by KMeans Ball Tree algorithm, ( k = 20 ) and the specific cluster of papers in which the query paper is in will be located. The system will take this cluster (has about 100 papers) as the candidate paper list. Second, the system will reduce the size of the candidates list to 50 by either (1) finding the top nearest neighbors to the centroid of the cluster, or (2) finding the top nearest neighbors to the query paper. Then in the third step, the Naive-Farthest algorithm will calculate all the candidate's cosine distance to the query paper on mechanism dimension and pick the top farthest 15 papers as final result. While the Farthest-Neighbor algorithm will set a random start in the top 50 papers and dynamically search for a subset of 15 papers that maximize the distance to each other. Cosine similarity is applied as the distance function used for Farthest Neighbor. The subset of 15 papers are papers that are near in purpose and diverse enough in mechanism. Both algorithms, unlike KNN-KMeans, can only support "near" on one aspect and "far" on another aspect.

**6.2 Performance and Analysis**

The Naive-Cosine-Distance algorithm is heavily based on cosine distance nearest neighbor algorithm. However the farthest papers on the "far" aspects found by cosine distance doesn't really mean the most "diverse" subset from "near" paper pool. We could say the two papers are similar or "near" on one aspects if their words embedding on that certain aspect are similar and close to each other, but it's really logically unclear and vague to say one paper is farther than another paper both compared to query seed just because the former one's embedding is "farther" than the latter's embedding to seed paper's embedding vector. Another important observation is that since our dataset currently only has 2071 labeled papers and for some query paper, the actual logically closely related papers judged by researchers might be fewer than 5 or 10 papers. So if we are trying to "diversify" the final results of ASE from this already limited true candidates pool from previous step, we will risk having very few papers relevant enough on the "near" aspect.

The KNN-KMeans algorithm works as expected to return highly related papers on "near" aspects but diverse on "far" aspects. However the absolute number of the results depends on how many candidates papers are removed. The final ASE result will vary from run to run because the KMeans clustering vary from run to run. One advantage of the KNN-KMeans "far" step is that it is still using the entire corpus information to define "far" since the KMeans is applied, clustering on the entire paper corpus. And the removal of papers in the same cluster from the candidates list won't destroy the order of the list. In other words, it won't destroy the "near" information we created from the first step, and thus shall guarantee that the diversification step of the algorithm won't make the final result list less relevant to the query paper on "near" aspects.

The "near" process of Naive-Farthest and Farthest-Neighbor algorithm(step 1, 2) works as expected to return topic related paper pools. For the "Far" diversification process, the Naive-Farthest faces the same problems the Naive Cosine Distance algorithm has. And in case of Farthest-Neighbor algorithm, we expect that step 3 will have various result papers from run to run, but the result is different from the expectation. After testing with 10 iterations on the fixed cluster for some query paper for several times, all the returned papers, except for the random start paper in step 3, are the same for one query paper (14 papers all appears as a result for each of the 10 iterations). One potential explanation for the situation is that the top 50 candidates papers are too sparsely distributed on the mechanism dimension ("far" aspect), and the marginal papers on this dimension are always the optimal choices for the step 3 farthest neighbors dynamical search, besides the random start paper. Consequently the performance of this algorithm might be limited on this small dataset. In situations that among the top 50 most near-purpose candidates, there are only 10% of the papers that are relevant to the query paper under human user's recognition , great chances are that the final diversified result list contains very few of those real "near" paper on purpose dimension. One more observation for these two algorithms is that, in step two if we choose to reduce the candidates pool by choosing the top nearest neighbors to the query paper, instead of to the centroid of the cluster from step one, the step one which uses KMeans for clustering are supposed to be skipped because we are not using the cluster information generated from step one.

Based on the observations above, the essence of the algorithms for current-stage ASE is that 1) search for top closely related papers from corpus to query paper on "near" aspects by K Nearest Neighbors, 2) apply a reasonable diversifying algorithm to pick a subset from the candidates list from previous step on the "far" aspects. To ensure the performance of the algorithm of ASE, currently for the Farthest-Neighbor algorithm, the key is to augment and expand the Solvent paper corpus to a much larger scale. If the dataset is large enough to guarantee that for each paper in the "near" paper candidates list, it is closely related to the the query paper from our database, no matter what research focus it is about. Then the diversification process on "far" aspects have greater potential to return interesting enough analogical search results for the user.

Given the limitation on our dataset size, we choose the KNN-KMeans algorithm as our ASE backend design for the experiment and evaluation, because such algorithm demonstrates best performance on our data and is much less sensitive to the limitation of Data size, compared to other algorithms for current stage. The diversification algorithm also applies the entire corpus for judgements which efficiently and completely utilized information of the vector embedding of "far" aspects.

# 7 User Evaluation and Results

We intend our system will be useful for anyone who wants to find out about what is going on at CMU in the field of AI. The intended users will include students, researchers, faculty and staff, alumni, media and companies. A prospective student may use the system to find out what is going on in the field of his/her research area. Students may use it to get new ideas for their research projects. Faculty researchers may use this system to keep track of new publications being published that may have some analogies with their own work. Media and Companies may use this system to track the progress being made in AI at CMU.

### 7.1 Experiment Setup and Results

To evaluate the performance of the Analogy Search, we choose Elasticsearch, a search engine based on Lucene library, as our baseline. Elasticsearch is a competitive and popular search engine used by Netflix, Stackoverflow, Medium and many other companies. With Elasticsearch as our baseline, we ran a small-scale user evaluation (5 users) for 21 query papers to perform A/B testings on Elasticsearch and Analogy Search. For analogy search, we retrieve all papers by near purpose, far mechanism. All the 5 users come from computer science-related master programs at CMU and have some background knowledge about the domain of the 21 selected query papers.

For each query paper, we randomly interleave the top 10 returned papers from Elasticsearch and Analogy Search(analogy search sometimes returns less than 10 papers. In that case, we interleave all its returned papers) . Then we ask the user to vote each returned paper as useful/maybe useful/not useful, and interesting/maybe interesting/not interesting. In total, we have collected 919 data points. Each data point contains information about the query paper, the returned paper, and their relationship. We can calculate how many of the returned paper is useful to the query paper, and how many of the returned paper is interesting with respect to the query paper and the user's comments.

### 7.2 Evaluation Metrics

For user evaluation, we performed A/B testing to compare the results of Elasticsearch and our Analogy Search in terms of two dimensions: usefulness and interesting. For each query paper , we randomly interleave the top 10 returned papers from Elasticsearch and Analogy Search and ask the user to vote each returned paper as useful/maybe useful/not useful, and interesting/maybe interesting/not interesting. We then aggregate the results to get the number of useful/maybe useful/not useful votes and the number of interesting/maybe interesting/not interesting votes for Elasticsearch and Analogy Search respectively. The evaluation metrics are the percentage of different votes and the percentage change of Analogy Search with respect to Elasticsearch, i.e. AnalogySearch_result - Elasticsearch_result ）/ Elasticsearch_result.

| | test_id | research_focus | seed_paper_name | SE | paper_id | paper_name | if_useful | useful_comment | if_interesting | interesting_comment |
|---|---------|----------------|-----------------|----|----------|------------|-----------|----------------|----------------|---------------------|
| 0 | 1 | NaN | Crowd Synthesis: Extracting Categories and Clu... | ES | wos_cs_69 | Learning Action Maps of Large Environments via... | maybe_useful | Similar problem addressed | maybe_interesting | The use of egocentric camera is interesting |
| 1 | 1 | NaN | Crowd Synthesis: Extracting Categories and Clu... | ES | msa_2071400413 | glance enabling rapid interactions with data u... | not_useful | not related | not_interesting | not related |
| 2 | 1 | NaN | Crowd Synthesis: Extracting Categories and Clu... | ES | msa_2009681737 | crowd synthesis extracting categories and clus... | useful | same | interesting | same |
| 3 | 1 | NaN | Crowd Synthesis: Extracting Categories and Clu... | ES | wos_cs_833 | Detecting Interesting Events Using Unsupervise... | not_useful | not related | not_interesting | not related |
| 4 | 1 | NaN | Crowd Synthesis: Extracting Categories and Clu... | ES | msa_2009681737 | crowd synthesis extracting categories and clus... | useful | same | interesting | same |

Figure 7: User Evaluation Data Points

Figure 8: A/B Testing Page

## 7. 2 Data Insights and Error Analysis

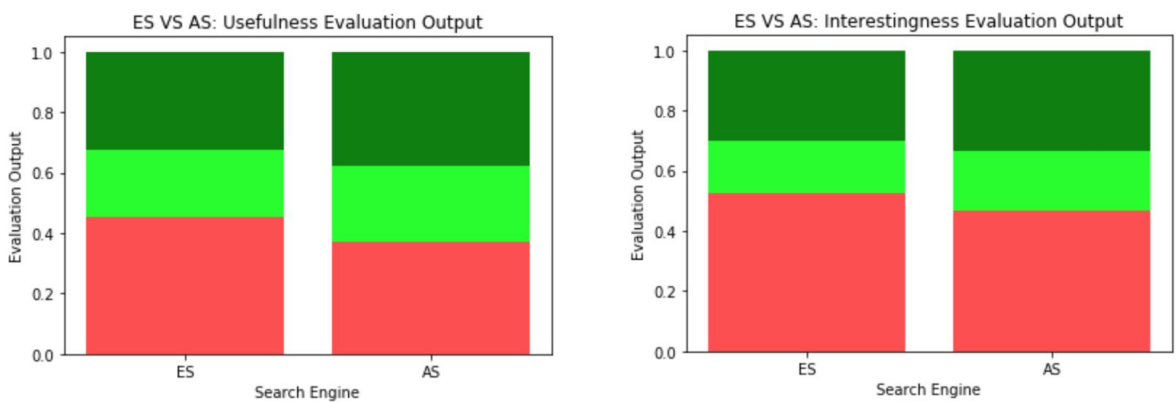

| | Usefulness Evaluation | | | | Interestingness Evaluation | | |
|---|---|---|---|---|---|---|---|
| | **ES** | **AS** | **Percentage Change** | | **ES** | **AS** | **Percentage Change** |
| **Useful** | 32.3 | 37.6 | +16% | **Interesting** | 30.3 | 33.2 | +9% |
| **Maybe Useful** | 22.3 | 25.3 | +13% | **Maybe Interesting** | 17.4 | 19.9 | +14% |
| **Not Useful** | 45.4 | 37.0 | -18% | **Not Interesting** | 52.3 | 46.8 | -10% |

From the above results, we can see that Analogy Search achieves around 10% better results than Elasticsearch in both usefulness(useful/maybe useful) and interestingness(interesting/maybe interesting) dimensions. This shows that Analogy Search provides significant improvements in helping our target users find more relevant and novel papers.

During the user evaluation, we observed that most users evaluated the returned papers by memorizing key words from the highlighted purpose and mechanism segments of the query paper. This may cause bias towards Analogy Search in that Analogy Search searches paper by purpose and mechanism segments while Elasticsearch searches by full abstracts. Also, we found that users did not have a uniform understanding of usefulness and interestingness of a paper. Under time pressure in evaluation, they tend to evaluate a paper's usefulness by its coverage of the query paper key words in their short-term memory.

Problems of data duplication and inconsistency also emerges during evaluation: papers with same title may have different external IDs and different purpose/mechanism segmentation. Also, since Elasticsearch does not search by segments, some papers returned by Elasticsearch may have no purpose/mechanism segments (no highlighted text) at all. This may indicate users that the described returned paper is from Elasticsearch.

## 8 Discussion

The essence of our analogical search engine algorithms is using KNN to pick the nearest neighbors to query paper from our corpus on the "near" aspect, and then figure out a reasonable and efficient algorithm to define the the process of finding "diverse" results on the "far" aspects. Because we are able to use the GloVe word embedding to compare similarity from query paper to the paper corpus on 4 categories of segments, instead of matching by full abstract embeddings, we are actually introducing external knowledge of the content and structure information of the abstract when doing the search. So during the search we are allowed to process the information of each functional segments separately with customized algorithms, thus focusing more on the crucial parts of the abstract information our users really cares by informing the system about their preference. These shall be the advantages that the ASE might outperform normal keyword matching search engine. But to improve the actual performance of ASE, more considerations and improvements should be applied on the customization on the "diverse" step of ASE. In a nutshell, we need to handle two conflicting objective, that is how to wisely implement diversity on the final result and on the same time, not lose the "near" information that the system collected from the "near" step. Furthermore, for the current algorithms like Farthest-Neighbor, the performance might be heavily dependent on the size of the dataset. A larger data corpus might significantly improve the performance of the diversity search and degree of relevance of the final result papers. Overall, we believe that the methods used in our Analogy Search Engine are better alternatives for searching any textual data which are written under certain writing standards, for example, grant proposals, product descriptions, wikipedia articles, etc.

## 9 Lessons Learned & Reflections

In the age of information overflow, efficiently getting the information one needs is crucial. Through this project, we learnt the difficulty of building good semantic search engines for textual data. We have experimented with numerous other academic search engines along the way. The deficiencies presented in these tools reflect some fundamentally unsolved problems in NLP, for example, how to represent different research intentions/purpose, and different research methods effectively. Embedding-based representation seem not working so well, while human annotated ontology classification might be too expensive to scale up. Nonetheless, we still firmly believe the methods used in ASE are a big improvement over current academic search engines. A more expansive and longer test will prove our ideas further.

## 10 Future Work

First we will develop a better and bigger dataset with each functional parts labeled. Joseph Chang and Xin Qian have done some work on building tag prediction models like Conditional Random Field model and LSTM model. However, with limited training set, the generalization ability of such models need to be evaluated before running on a larger dataset.

Second, we will rethink the embedding process, especially since we are analyzing information-rich textual information, paper abstracts. Many acronyms like LSTM, AutoML, CNN needs more context for the GloVe word embedding to work. Ideally, we would want such embedding techniques to accurately capture the "farness" and "closeness" of research purpose/intentions, research backgrounds and research methods.

Third, we will release a beta website for our Analogy Search soon so researchers can spend more time using our system instead of sitting down with us for hours to go through the long list of returned results. During the user evaluation studies, we found that researchers need longer time to think about the returned paper abstracts. Inspiration does not happen as soon as the researchers finish reading the abstract. This reflects an important difference between commodity recommender system and academic paper recommender system. Since academic research paper abstracts are very condensed information instruments which require a much longer time to think about than common commodities.

## 11 Conclusion

In this project, we developed and tested an innovative search engine for research paper abstracts. We showed that Analogy Search Engine outperforms baseline information retrieval elasticsearch on 2000 CMU AI papers. Researchers are more likely to find useful and interesting results with ASE. We also found that close cosine distance of the embeddings strongly suggests similarly in contents, while far cosine distance suggests difference in contents, but not enough to demonstrate far in meanings (different methods, different findings). Furthermore, we observed that researchers judge content similarity / dissimilarity by condensing sentences into keywords. Providing keywords may help researchers appreciate more the similarity and diversity in search results.

## 12 Acknowledgments

We want to thank Professor Niki Kittur at Human Computer Interaction Institute of CMU and Professor Joel Chan at iSchool of UMD for giving us the project opportunity and all the great mentoring along the course of the project. Joseph Chang, Xin Qian, Hyeonsu Kang also provide us many constructive supports for our project. We also want to thank Dhruv Shah, Baoyu Jing, Tian Jin, etc. for their efforts in evaluating our Analogy Search Engine.